\documentclass[usenatbib, useAMS]{mnras}

\usepackage{graphicx}
\usepackage{amsmath,paralist,xcolor,xspace,amssymb}
\usepackage{times}

\newcommand{\msun}{{\rm{M}_{\sun}}}
\newcommand{\rsun}{{r_{\sun}}}

\newcommand{\lcdm}{$\Lambda$CDM\xspace}

\newcommand{\eagle}{\textsc{eagle}\xspace}
\newcommand{\apostle}{\textsc{apostle}\xspace}
\newcommand{\dmo}{\textsc{dmo}\xspace}

\newcommand{\gadget}{\textsc{gadget}\xspace} 
\newcommand{\subfind}{\textsc{subfind}\xspace}
\newcommand{\anarchy}{\textsc{anarchy}\xspace}

\newcommand{\degree}{\degr}

\title[DM annihilation signal from simulated MWs]{Dark matter annihilation
  radiation in hydrodynamic simulations of Milky Way haloes}

\author[M. Schaller et al.]  {Matthieu Schaller$^1$\thanks{E-mail: matthieu.schaller@durham.ac.uk},
                              Carlos S. Frenk$^1$,
                              Tom Theuns$^1$,
                              Francesca Calore$^2$,  \newauthor
                              Gianfranco Bertone$^2$,
                              Nassim Bozorgnia$^2$,
                              Robert A. Crain$^3$,
                              Azadeh Fattahi$^4$, \newauthor
                              Julio F. Navarro$^{4,5}$,
                              Till Sawala$^1$ \&
                              Joop Schaye$^6$
                              \\
$^1$Institute for Computational Cosmology, Durham University, South Road,
                              Durham, UK, DH1 3LE\\                              
$^2$GRAPPA, University of Amsterdam, Science Park 904, 1090 GL Amsterdam, The
                              Netherlands\\ 
$^3$Astrophysics Research Institute, Liverpool John Moores University, 146
                              Brownlow Hill, Liverpool L3 5RF, UK\\ 
$^4$Department of Physics and Astronomy, University of Victoria, 3800 Finnerty
                              Road, Victoria, British Columbia V8P 5C2, Canada\\
$^5$Senior CIfAR Fellow\\
$^6$Leiden Observatory, Leiden University, Postbus 9513, 2300 RA Leiden, The
                              Netherlands 
}
\begin{document}

\date{\today}

\pagerange{\pageref{firstpage}--\pageref{lastpage}} \pubyear{2015}

\maketitle

\label{firstpage}

\begin{abstract}
We obtain predictions for the properties of cold dark matter
annihilation radiation using high resolution hydrodynamic zoom-in
cosmological simulations of Milky Way-like galaxies (\apostle project)
carried out as part of the ``Evolution and Assembly of GaLaxies and
their Environments'' (\eagle) programme. Galactic halos in the
simulation have significantly different properties from those assumed
in the ``standard halo model'' often used in dark matter detection
studies. The formation of the galaxy causes a contraction of the dark
matter halo, whose density profile develops a steeper slope than the
Navarro-Frenk-White (NFW) profile between $r\approx1.5~\rm{kpc}$ and
$r\approx10~\rm{kpc}$. At smaller radii, $r\lesssim1.5~\rm{kpc}$, the
halos develop a flatter than NFW slope.  This unexpected feature may
be specific to our particular choice of subgrid physics model but
nevertheless the dark matter density profiles agree within $30\%$ as
the mass resolution is increased by a factor $150$. The inner regions
of the halos are almost perfectly spherical (axis ratios $b/a > 0.97$
within $r=1~\rm{kpc}$) and there is no offset larger than $45~\rm{pc}$
between the centre of the stellar distribution and the centre of the
dark halo. The morphology of the predicted dark matter annihilation
radiation signal is in broad agreement with $\gamma$-ray observations
at large Galactic latitudes ($b\gtrsim3^\circ$). At smaller angles,
the inferred signal in one of our four galaxies is similar to that
which is observed but it is significantly weaker in the other three.
\end{abstract}


\section{Introduction}
\label{sec:introduction}

Uncovering the unknown nature of the dark matter is one of the greatest
challenges of modern cosmology and particle physics. Since the original
suggestion that the dark matter might consist of massive, cold, weakly
interactive, neutral particles, a large body of empirical evidence has
consolidated this hypothesis which, however, can only be confirmed by the
detection of the particles themselves.  Among the possible candidates,
supersymmetric particles \citep[see][for a review]{Jungman1996} such as
neutralinos are attractive options that current particle accelerator experiments
might detect.

An interesting property of many particle candidates for cold dark matter is that
they annihilate into standard model particles, including photons. This opens up
the exciting possibility of attempting to detect such photons from space. The
requirement that weakly interacting particles provide the measured dark matter
density in the Universe today suggests a plausible particle mass range of order
$m_\chi = 10 - 1000~\rm{GeV}$, leading to the emission of $\gamma$-ray photons
in or below that energy range when two dark matter particles annihilate
\citep[see review by][]{Bertone2005}.

The Large Area Telescope aboard the Gamma-Ray Space Telescope (\emph{Fermi})
\citep{GLAST1999} has, over the last few years, produced the most detailed maps
of the $\gamma$-ray sky, covering a large energy range
($20~\rm{MeV}-500~\rm{GeV}$) with a resolution of a few arcmin at the highest
energy end of the spectrum. Analysing the Fermi data around the Galactic Centre
(GC), a number of authors
\citep{Goodenough2009,Hooper2010,Hooper2011,Abazajian2012,Gordon2013,
  Hooper2013,Macias2014,Daylan2014,Abazajian2014,Calore2015a} have claimed the
detection of extended diffuse excess emission above the other known
astrophysical sources. This excess emission, peaking at $E\approx2~\rm{GeV}$,
was found to be broadly consistent with the expected signal from dark matter
annihilation. In particular, the flux decreases with distance from the GC as
$r^{-\Gamma}$ with slope, $\Gamma=2.2-2.4$, only slightly steeper than the
asymptotic inner slope, $\Gamma=2$, of flux originating from the NFW
\citep{Navarro1996a,Navarro1997} density profiles found in dark matter only
simulations of halos.

However, potential systematic effects in the analysis of the $\gamma$-ray data
could be introduced by incorrect point source subtraction or the modelling of
the diffuse backgrounds \citep{Fermi2012}.  Alongside the dark matter
interpretation of the Galactic Centre excess, astrophysical explanations have
been proposed. For example, a population of as yet unresolved millisecond
pulsars (MSP) \citep{Abazajian2011,Gordon2013,Yuan2014,Cholis2014} or young
pulsars \citep{OLeary2015} associated with the bulge\footnote{The thick disk
  population of MSPs and pulsars are unlikely to contribute more than 10\% to
  the GeV excess \citep{Calore2014b}.} of the Milky Way. In contrast to the
conclusions of \cite{Hooper2013b} and \cite{Cholis2014}, \cite{Bartels2015} and
\cite{Lee2015} have argued that the excess could be due to such a pulsar
population.  Alternatively, \cite{Linden2012, Carlson2014,
  Petrovic2014} and \cite{Cholis2015} have
suggested that the excess could originate from the injection of very high energy
cosmic rays during past activity in the Galactic Centre.

Besides the spectral shape, another property that can help distinguish the
potential sources of $\gamma$-rays contributing to the excess is the morphology
of the signal. A dark matter origin would require the excess to extend over tens
of degrees on the sky \citep{Springel2008b,Serpico2008,Nezri2012}. An excess
with the same spectral shape extending over a large angular range, with
emissivity decreasing with distance from the GC, would strengthen the
interpretation of the excess as originating from dark matter annihilation. Using
multiple regions between $2\degree$ and $20\degree$ from the GC and a large
range of Galactic diffuse emission (GDE) models, \cite{Calore2015a} found that
the excess emission is consistent with a dark matter particle of mass
$m_\chi=49^{+6.4}_{-5.4}~\rm{GeV}$ annihilating into a $b\bar b$ quark pair then
producing photons and is distributed following a generalised NFW profile (gNFW,
see equation \ref{eq:gNFW} below) with a slope $\gamma=1.26\pm0.15$. A similar
spatial distribution was found by \cite{Daylan2014} who suggested a slope for
the inner profile in the range $\gamma=1.1-1.3$. With the increasing precision
of these measurements and of the foreground modelling, it has become crucial to
refine the theoretical models for the distribution of dark matter at the centre
of galaxies in a $\Lambda$CDM context. Characterising the dark matter profile
slope and sphericity as well as investigating the potential offset between the
dark matter and the GC are all important tasks for theorists.

Work based on dark matter only simulations has shown that the dark matter is
distributed following an NFW density profile with a scale length, $r_{\rm s}$,
which varies with halo mass \citep[e.g.][]{Navarro1997,
  Neto2007,Duffy2008,Dutton2014}. Higher-resolution simulations
\citep{Springel2008a} have shown that the very inner parts of dark matter
profiles might be slightly shallower than the asymptotic NFW slope of $\gamma=1$
\citep{Navarro2010}. Similarly, predictions for the signal coming from subhaloes
have also been made using these simulations \citep{Kuhlen2008,Springel2008b},
effectively proposing a test of the cold dark matter paradigm. At the other end
of the halo mass range, \cite{Gao2012} argued that nearby rich clusters
provide a signal with a higher signal-to-background ratio than the
Milky Way's satellites. Thus far observational measurements have proved
inconclusive and the only claimed detection comes from the centre of our own
Milky Way, where precise predictions from dark matter simulations have been made
\citep{Springel2008b}.

However, these studies all ignored the effects of the forming galaxy on the
structure of the dark matter halo. Mechanisms such as dark matter contraction
\citep[e.g.][]{Barnes1984, Blumenthal1986,Gnedin2004} can drag dark matter
towards the centre, steepening the profile. Conversely, perturbations to the
potential, due for instance to feedback from stars or supermassive black holes
or the formation of a bar, can lead to a flattening of the very central regions
\citep[e.g.][]{Navarro1996b, Weinberg2002,
  Mashchenko2006,PontzenGovernato2012}. The correct balance between these
mechanisms can only be understood by performing high-resolution hydrodynamic
simulations of Milky Way-like galaxies using a physical model validated by
comparison with a wide range of other observables.

In this study we use two ``zoom'' simulations of Local Group environments
\citep[the \apostle project:][]{Sawala2014, Fattahi2015} performed within the
framework of the ``Evolution and Assembly of GaLaxies and their Environments''
(\eagle) suite \citep{Schaye2015,Crain2015}. These simulations have been shown
to reproduce a large number of observables of the galaxy population at low and
high redshifts, as well as the satellite galaxy luminosity functions of the
Milky Way and Andromeda galaxies with unprecedented
accuracy. \cite{Schaller2014} showed that the \eagle simulations produce
galaxies with rotation curves that are in unprecedented agreement with
observations of field galaxies, suggesting that the matter distribution in the
simulated galaxies is realistic and that the main that effects of baryons on
halos are accurately captured by the simulations. Note, however, that
\cite{Calore2015c} showed that the goodness of fit of the simulated data to the
observed Milky Way rotation curve is lower in the highest resolution zoomed-in
simulations. The \eagle simulations therefore provide an excellent test-bed for
the interpretation and analysis of the Fermi excess.

This paper is structured as follows. In Section \ref{sec:simulations} we
introduce the simulation setup used. We investigate the dark matter density
profiles of our halos in Section \ref{sec:profiles} and analyse the dependence
of the annihilation signal on the angle with respect to the Galactic Centre in
Section \ref{sec:signal}. We summarise our findings and conclude in Section
\ref{sec:summary}.

Throughout this paper, we assume a \emph{WMAP7} flat \lcdm cosmology
\citep{WMAP7} ($h = 0.704$, $\Omega_b = 0.0455$, $\Omega_m = 0.272$
and $\sigma_8= 0.81$), express all quantities without $h$ factors and
assume a distance from the GC to the Sun of $\rsun=8.5~\rm{kpc}$.

\section{The simulations}
\label{sec:simulations}

The simulations used in this study are based on the \eagle simulation code
\citep{Schaye2015,Crain2015} . We summarize here the parts of model relevant to
our discussion.

\subsection{Simulation code and subgrid models}

The \eagle code is based on a substantially modified version of the \gadget
code, last described by \cite{Springel2005}. The modifications include the use
of a state-of-the-art implementation of Smoothed Particle Hydrodynamics (SPH),
called \anarchy \citep[Dalla Vecchia (in prep.), see also][]{Schaller2015b},
based on the pressure-entropy formulation of SPH
\citep{Hopkins2013}. Gravitational interactions are computed using a Tree-PM
scheme.

The cooling of gas and its interaction with the background radiation is
implemented following the recipe of \cite{Wiersma2009a} who tabulated
photoheating and cooling rates element-by-element (for the 11 most important
elements) in the presence of the UV and X-ray backgrounds inferred by
\cite{Haardt2001}.  To prevent artificial fragmentation, a pressure floor in the
form of an effective equation of state, $P_{\rm eos} \propto \rho^{4/3}$,
designed to mimic the mixture of phases in the interstellar medium (ISM)
\citep{Schaye2008}, is applied to the cold and dense gas. Star formation is
implemented using a pressure-dependent prescription that by construction
reproduces the observed Kennicutt-Schmidt star formation law \citep{Schaye2008}
and uses a density threshold that captures the metallicity dependence of the
transition from the warm, atomic to the cold, molecular gas phase
\citep{Schaye2004}.  Star particles are treated as single stellar populations
with a \cite{Chabrier2003} initial mass function (IMF) evolving along the tracks
advocated by \cite{Portinari1998}. Metals from supernovae (SNe) and AGB stars are
injected into the ISM following the model of \cite{Wiersma2009b} and stellar
feedback is implemented via the stochastic injection of thermal energy into the
gas neighbouring newly-formed star particles as described by
\cite{DallaVecchia2012}. Galactic winds hence form naturally without having to
impose a direction, velocity or mass loading factor. The amount of energy
injected into the ISM per feedback event is dependent on the local gas
metallicity and density in an attempt to take into account the unresolved
structure of the ISM \citep{Schaye2015, Crain2015}.  Supermassive black hole
seeds are injected in halos above $10^{10}h^{-1}\msun$ and grow through mergers
and the accretion of low angular momentum gas
\citep{RosasGuevara2013,Schaye2015}. AGN feedback is modelled by stochastically
injecting thermal energy into the gas directly surrounding the black hole
\citep{Booth2009,DallaVecchia2012}. Halos are identified using the FOF algorithm
\citep{Davis1985} and substructures within them are identified in
post-processing using the \subfind code \citep{Dolag2009}.

The subgrid model was calibrated (by adjusting the efficiency of stellar
feedback and the accretion rate onto black holes) so as to reproduce the present
day stellar mass function and galaxy sizes, as well as the relation between
galaxy stellar masses and supermassive black hole masses \citep{Schaye2015,Crain2015}. As
argued by \cite{Schaye2015} (Sec.~2), numerical convergence in the traditional sense
(\emph{strong} convergence) cannot be achieved when new physical processes are
resolved at each resolution. In this case, the free parameters of the model
should be re-calibrated to match the same pre-defined set of observables
(\emph{weak} convergence). This can be done in cosmological simulations of
representative periodic volumes \citep{Crain2015}, but it is much more difficult
to achieve for ``zoom-in'' simulations of a few objects. In this case, even weak
convergence is difficult to establish (See Sec.~\ref{sec:convergence}).

\subsection{Selection of Milky Way halos}

The two volumes used in this work are zoom resimulations of regions extracted
from a dark matter only simulation of $100^3~\rm{Mpc}^3$ with $1620^3$
particles. The halos were selected to match the observed dynamical constraints
of the Local Group \citep[\apostle project:][]{Sawala2014,Fattahi2015}. Each
volume contains a pair of halos in the mass range $M_{200}=5\times10^{11}\msun$
to $M_{200}=2.5\times10^{12}\msun$ that will host analogues of the Milky Way and
M31. We use the two halos in volumes AP-1 and AP-4 of the \apostle suite
\citep[see Table 2 of][where other relevant data are listed]{Fattahi2015}. The high-resolution region encloses a sphere
larger than $2.5~\rm{Mpc}$ around the centre of mass of the two halos at
$z=0$. The dark matter particle mass in the zoom regions was set to
$5\times10^4\msun$, whilst the primordial gas particle mass was set to
$1\times10^4\msun$. The initial conditions were generated from $\Lambda$CDM
power spectra using 2\textsuperscript{nd} order Lagrangian perturbation theory
\citep{Jenkins2010} and linear phases were taken from the Gaussian white noise
field \textsc{panphasia} \citep{Jenkins2013}. The gravitational softening length
was set to $\epsilon=134~\rm{pc}$ (Plummer equivalent) at $z<2.8$ and was kept
fixed in comoving units at higher redshifts.  Simulations without baryonic
components were run with the exact same setup and are labelled as \dmo in what
follows.

\section{Dark matter distribution in the centre of the halos}
\label{sec:profiles}

In this section, we analyse the dark matter distribution of the haloes of
simulated Milky Way galaxies. We consider both central galaxies in each of the
two simulation volumes as Milky Way-like galaxies.

\subsection{Profiles without baryon effects}
\label{ssec:profilesDMO}

The analysis of the GC excess is often performed using an assumed analytic
density profile shape for the dark matter. This profile is a generalisation of
the NFW profile for which the asymptotic inner slope is a free parameter
$\gamma$:

\begin{equation}
 \rho_{\rm DM}(r) = 
\frac{\rho_{\rm s}}{\left(r/r_{\rm s}\right)^{\gamma}\left(1+r/r_{\rm s}\right)^{
3-\gamma}}
\label{eq:gNFW}
\end{equation}
The NFW profile is recovered for $\gamma=1$. This generalized form of the
density profile is not supported by numerical simulations of collisionless dark
matter \citep[e.g.][]{Navarro2010} but is a useful way to parametrise the
deviation from the NFW shape in the very centres of halos as a result of
baryonic effects. As the measurements of the GC excess only span a range of a
few kiloparsecs, the value of the scale radius $r_{\rm s}$ cannot be constrained
observationally and is typically fixed to $r_{\rm s}=20~\rm{kpc}$, in broad
agreement with simulation results for Milky Way-like halos
\citep[e.g.][]{Neto2007, Dutton2014}. The normalisation of the profile,
$\rho_{\rm s}$, is degenerate with other particle physics parameters (see
Section \ref{sec:signal}) and is usually fixed by requesting that the density of
dark matter at the location of the Sun\footnote{Note that for simplicity we use
  units convenient for particle physics applications. Units more friendly to
  astronomers are recovered using the conversion
  $1~\msun\cdot\rm{kpc}^{-3}=3.795 \times 10^{-8}~\rm{GeV}\cdot\rm{cm}^{-3}$.}
is $\rho_{\rm DM}(\rsun)=0.4~\rm{GeV}\cdot\rm{cm}^{-3}$, in agreement with local
dynamical constraints \citep{Catena2010, Iocco2011}.

\begin{figure}
\includegraphics[width=1.\columnwidth]{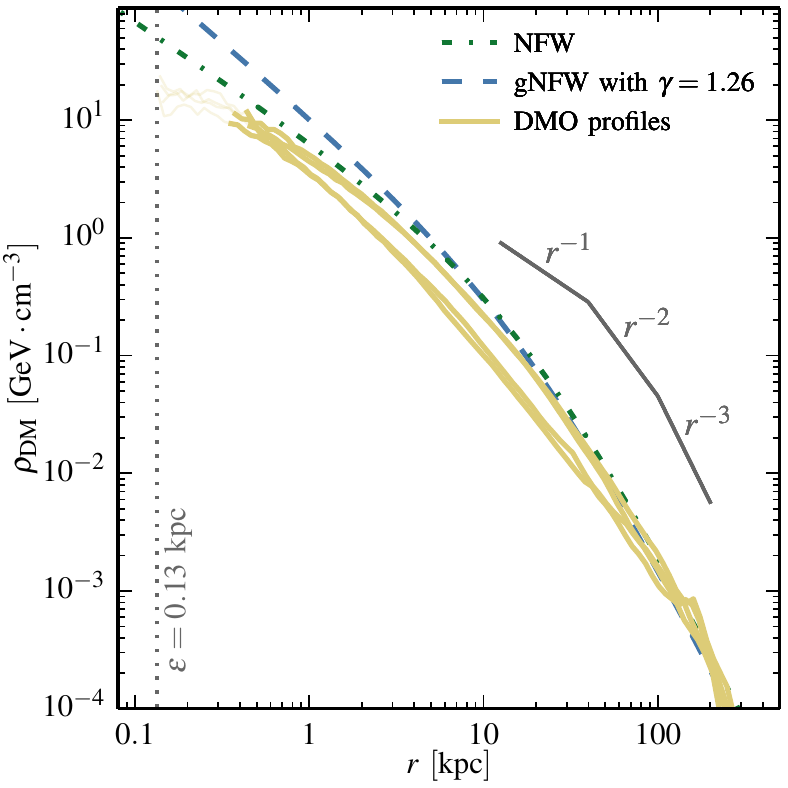}
\caption{The dark matter density profiles of the four halos in the simulations
  without baryons (yellow solid lines). Thinner lines are used at radii smaller
  than the convergence radius of the simulation. The vertical dotted line
  indicates the simulation's gravitational softening length. The green
  dot-dashed and blue dashed lines correspond to an NFW and gNFW with
  $\gamma=1.26$ profiles respectively, both normalised to $\rho(\rsun)=0.4~{\rm
    GeV\cdot cm^{-3}}$ and with a scale-length $r_{\rm s}=20~\rm{kpc}$. As
  expected, the simulated profiles display a shape similar to the plotted NFW
  model, but with a lower normalisation than the standard halos.}
\label{fig:DMOprofiles}
\end{figure}

In order to quantify the effects of baryons on the dark matter distribution, it
is worth first considering the profiles extracted from the simulations without
baryonic physics. In Fig.~\ref{fig:DMOprofiles}, we show the dark matter density
profiles of our four halos. Thick lines are used at radii greater than the
resolution limit ($R_{\rm P03}\approx350-450~\rm{pc}$ depending on the halo) set
by the criterion of \cite{Power2003} and thin lines are used at smaller
radii. The softening length is indicated by the vertical dotted line. The green
dot-dashed and blue dashed lines correspond to NFW and gNFW with $\gamma=1.26$
(the best-fitting value of \cite{Calore2015a} to the excess) profiles
respectively, both normalised, as discussed above, to $\rho_{\rm
  DM}(\rsun)=0.4~{\rm GeV\cdot cm^{-3}}$ and with a scalelength $r_{\rm
  s}=20~\rm{kpc}$. As expected, the profiles are in good agreement with the NFW
model albeit with a lower normalisation. The best-fitting NFW profile parameters
to our halos are given in Table \ref{tab:halos_DMO}. The usual choice of $r_{\rm
  s}=20~\rm{kpc}$ is in good agreement with our simulated halos but the
normalisation of our halos is lower than what is often assumed in the
literature. When baryon effects are neglected, an inner slope close to
$\gamma=1.26$ is clearly incompatible with the simulations.

\begin{table}
  \caption{Properties of the four simulated \dmo halos
    (Fig. \ref{fig:DMOprofiles}) and the best-fitting NFW parameters to their
    dark matter density profiles. The spherical overdensity masses and radii are
    given with respect to the critical density of the universe.}
\label{tab:halos_DMO}
\begin{center}
\begin{tabular}{|c|c|c|c|c|c|}
Halo & $M_{200}$ & $R_{200}$ & $R_{\rm P03}$ &
    $r_{\rm s}$ & $\rho_{\rm DM}(\rsun)$\\ 
& $[\msun]$ & $[\rm{kpc}]$ & $[\rm{pc}]$ & $[\rm{kpc}]$ & $[\rm{GeV}\cdot\rm{cm}^{-3}]$\\
\hline 
$1$ & $1.65\times10^{12}$ & $243.2$ & $435$ &  $22.4$ & $0.290$ \\
$2$ & $1.09\times10^{12}$ & $212.0$ & $445$ &  $20.1$ & $0.132$ \\
$3$ & $1.35\times10^{12}$ & $226.9$ & $344$ &  $23.2$ & $0.162$ \\
$4$ & $1.39\times10^{12}$ & $229.4$ & $358$ &  $19.8$ & $0.281$ \\
\end{tabular}
\end{center}
\end{table}

\subsection{Profiles in the simulations with baryons}
\label{ssec:profiles}

\begin{figure}
\includegraphics[width=1.\columnwidth]{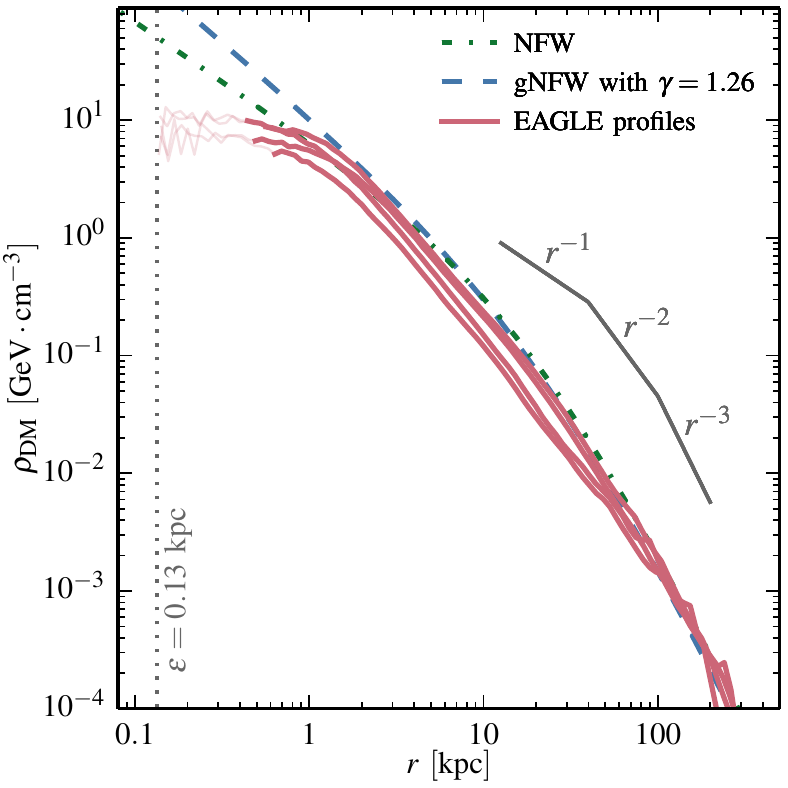}
\caption{The dark matter density profiles of the four halos in the simulations
  with baryons physics (red solid lines). Thinner lines are used at radii
  smaller than the convergence radius of the simulation. The vertical dotted
  line indicates the simulation's gravitational softening length. The green
  dot-dashed and blue dashed lines correspond to an NFW and gNFW with
  $\gamma=1.26$ profiles respectively, both normalised to $\rho(\rsun)=0.4~{\rm
    GeV\cdot cm^{-3}}$ and with a scalelength $r_{\rm s}=20~\rm{kpc}$. The
  profiles display a logarithmic slope steeper than $-1$ between
  $r\approx1.5~\rm{kpc}$ and $r\approx10~\rm{kpc}$, in broad agreement with the
  profiles inferred from observations by \citet{Calore2015a}. At radii
  $r\leq1~\rm{kpc}$ the profile is significantly shallower than NFW profiles
  are.}
\label{fig:profile}
\end{figure}

We now turn towards the dark matter profiles in the simulations
including baryons. In Fig.~\ref{fig:profile}, we show the dark matter
density profiles of the four halos simulated with the full baryonic
model. As for the previous figure, the lines are thin at radii less
than the convergence radius of the simulation $R_{\rm P03}$ and the
dashed lines correspond to the NFW and gNFW profiles normalised to
$\rho(\rsun)=0.4~{\rm GeV\cdot cm^{-3}}$. The simulated profiles
present two interesting features when compared to the \dmo results. In
the range $1.5-10~\rm{kpc}$, the profile is significantly steeper than
an NFW profile and at radii $r\lesssim1~\rm{kpc}$, the profiles
display a significant flattening. Our profiles thus display a
combination of dark matter contraction and a flattening further in.
The properties of the dark matter distribution in the central regions
are, however, particularly sensitive to the choice of subgrid model so
these results, particularly the flattening of the profile, should be
regarded as tentative and, by no means, as a generic prediction of
\lcdm.

These halos are clearly not well described over their entire radial range by
any of the profiles commonly found in the literature. The main properties of the
halos are given in Table \ref{tab:halos}. Note that in agreement with the
findings of \cite{Schaller2014} for the lower-resolution periodic \eagle volume, the
halo masses $M_{200}$ (and hence radii $R_{200}$) are lower than in the
simulation that did not include baryon physics. A consequence of the steepening
of the profile due to contraction is the slight increase in the local dark
matter density $\rho_{\rm DM}(\rsun)$ (column 6 of the tables), which, however,
remains lower than the commonly adopted value of $0.4~\rm{GeV}\cdot\rm{cm}^{-3}$
in each case. Clearly, the simulated profiles will not be well described by a
gNFW profile at radii $r\lesssim1.5~\rm{kpc}$. It is, however, instructive to
find the best-fitting profile at larger radii for comparison with the models
used to characterise the Fermi excess. The best-fitting asymptotic slopes are
given in column 5 of table \ref{tab:halos}. For all four halos, the slopes are
steeper than the value ($\gamma=1.26\pm0.15$) found by \cite{Calore2015a} when
modelling the GC excess. We note, however, that the simulated profiles are in
broad agreement with the gNFW profile of \cite{Calore2015a} (blue dashed line),
if the overall normalisation is, once again, ignored. The baryons
significantly steepen the profiles at radii $r\gtrsim1.5~\rm{kpc}$.

\begin{table}
  \caption{Properties of the four simulated \eagle halos
    (Fig. \ref{fig:profile}) and the best-fitting gNFW asymptotic slope
    $\gamma$ to their dark matter density profiles in the radial range
    $r>1.5~\rm{kpc}$.}
\label{tab:halos}
\begin{center}
\begin{tabular}{|c|c|c|c|c|c|}
Halo & $M_{200}$ & $R_{200}$ & $R_{\rm P03}$ &  $\gamma$ & $\rho_{\rm DM}(\rsun)$\\ 
& $[\msun]$ & $[\rm{kpc}]$ & $[\rm{pc}]$ & $[-]$ & $[\rm{GeV}\cdot\rm{cm}^{-3}]$\\
\hline 
$1$ & $1.56\times10^{12}$ & $238.8$ & $559$ & $1.38$  & $0.310$ \\
$2$ & $1.01\times10^{12}$ & $206.8$ & $592$ & $1.47$  & $0.160$ \\
$3$ & $1.12\times10^{12}$ & $213.7$ & $438$ & $1.73$  & $0.204$ \\
$4$ & $1.16\times10^{12}$ & $216.2$ & $462$ & $1.49$  & $0.280$ \\
\end{tabular}
\end{center}
\end{table}

At radii $r\lesssim1.5~\rm{kpc}$, the density profiles deviate
significantly from the cusp seen in the \dmo simulation. At the
resolution limit, $R_{\rm P03}=450-600~\rm{pc}$, the simulated
profiles exhibit a density between $2.5$ and $4.2$ times lower than
the best-fit gNFW profile inferred from observations. This flattening
is an important feature since the densest regions of the haloes
dominate the $\gamma$-ray emission. No such flattening was seen in the
reference \eagle simulations, which, however, had over 200 times
poorer mass resolution than the Local Group \apostle simulations used
here. Here, the flattening is well resolved since it occurs at radii
significantly larger than the \cite{Power2003} convergence radius.
This indicates that the flattening is not a result of poor sampling
but rather a consequence of our specific choice of subgrid model.

At high redshift all four examples had developed the cuspy central profile that
is characteristic of dark matter haloes. However, the inner profiles flattened
during events that are clearly associated with violent star formation in the
inner few kiloparsecs. In one example, the cusp regenerated before being
flattened again by a new episode of violent star formation activity. This
phenomenon is reminiscent of the cusp-destroying mechanism proposed by
\cite{Navarro1996b} in which the sudden removal of dense, self-gravitating gas
from the centre by a starburst redistributes the binding energy of the central
regions.  Related processes have been seen in recent simulations, mostly of
dwarf galaxies \citep[e.g.][]{Governato2010, Duffy2010,
  Maccio2012,PontzenGovernato2012,Teyssier2013,Onorbe2015}.  We defer a detailed
discussion of the causes of this interesting phenomenon to a separate study.

\subsection{Resolution and convergence considerations}
\label{sec:convergence}

In the previous two subsections, we used the criterion of \cite{Power2003} as
the resolution limit of our simulations. This criterion, based on the two-body
relaxation timescale, was derived using purely collisionless simulations and was
designed to ensure that the enclosed mass at a given radius, $R_{\rm P03}$, is
within $10\%$ of the value obtained using a higher resolution simulation. In the
cases where baryonic effects are simulated, it is unclear whether this criterion
still applies and even whether numerical convergence in the usual sense can be
achieved (see discussion in \cite{Schaye2015}). \cite{Schaller2014} demonstrated
that stacked haloes, extracted from the \eagle volumes, are well converged when
using this simple criterion but it is unclear whether this remains true when
individual haloes are considered. Of particular concern is the use of the pressure
floor for the densest gas, which sets an artificial scale below which gas cannot
be compressed. For our simulations, at all resolutions, this pressure floor
ensures that Jeans lengths above $\lambda_{\rm J} \approx750~\rm{pc}$ can be
resolved and prevents the collapse of gas clouds of smaller sizes. It is
therefore possible that the profiles may be modified by this pressure floor at
radii $r\sim\lambda_{\rm J}$, which, incidentally, is similar to the value of
$R_{\rm P03}$ in our highest resolution simulation.

\begin{figure}
\includegraphics[width=1.\columnwidth]{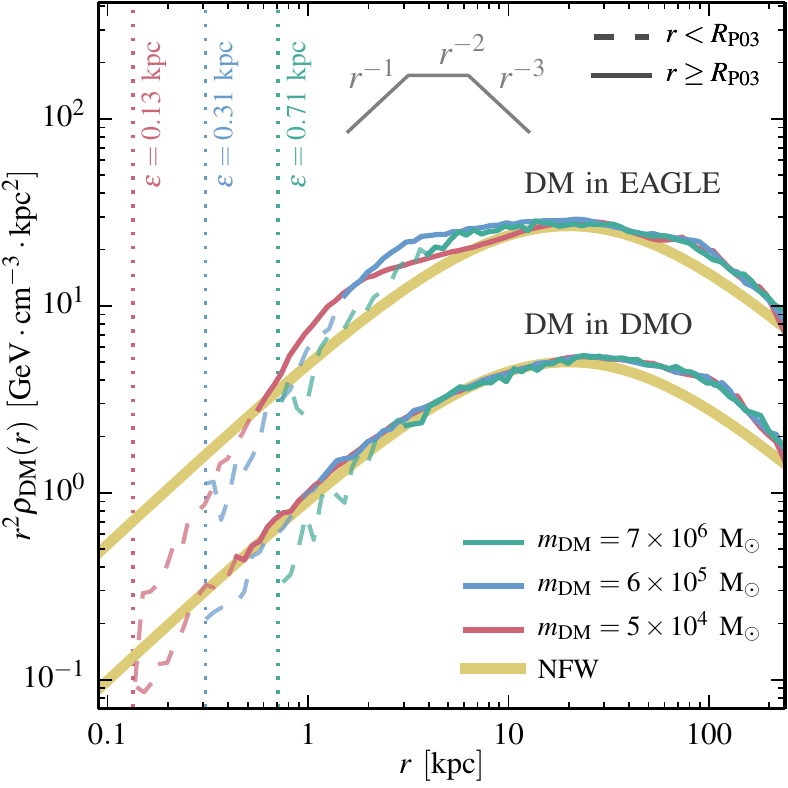}
\caption{Convergence test for the dark matter profiles of one of the haloes
  (Halo 1) in both the simulation with baryons (upper set of lines) and without
  baryons (lower set of lines, rescaled by a factor of $0.2$ for clarity). The
  green, blue and red lines correspond to the simulations with a dark matter
  particle mass of $7\times10^6~\msun $, $6\times10^5~\msun $ and
  $5\times10^4~\msun$ respectively. Dashed lines are used at radii smaller than
  the convergence radius, $R_{\rm P03}$, of each resolution. The vertical dotted
  lines indicate the softening length of each simulation. The thick yellow lines
  show an NFW profile with $r_{\rm s}=20~\rm{kpc}$. The profiles in the \dmo
  simulation are well converged even at $r< R_{\rm P03}$. In the \eagle
  simulation, the profiles display a lower level of convergence but nevertheless
  agree within $30\%$ at $r> R_{\rm P03}$. The differences between the various
  resolutions are, however, smaller than the difference relative to the \dmo
  case. The behaviour of this particular halo is typical of the four cases we
  simulated. These are all shown in Appendix \ref{sec:allConvergence}.}
\label{fig:convergence}
\end{figure}

In Fig.~\ref{fig:convergence} we show the density profiles (multiplied by $r^2$
to reduce the dynamic range) of the dark matter extracted from one of our
simulations run at three different resolutions, separated by factors of $12$ in
particle mass. The bottom set of curves are extracted from the simulation
without baryons (and have been multiplied by a factor $0.2$ for clarity), whilst the
upper set of lines are taken from the simulations with baryons. Dashed lines are
used at radii $r<R_{\rm P03}$ and the vertical dotted lines indicate the
softening lengths in each of the three simulations. As a guide, NFW
profiles with $r_s=20~\rm{kpc}$ and the same value of $\rho_{\rm DM}(\rsun)$ as
the highest resolution simulation are shown using yellow lines. Similar figures for
the three other haloes are shown in Appendix \ref{sec:allConvergence}.

In the \dmo simulations (lower set of curves), the profiles of different
resolution converge very well and, as expected, the criterion of
\cite{Power2003} (which refers to the enclosed mass) provides a good, conservative,
estimate of the radius above which the \emph{density} profiles are converged. In
fact, the density profiles are converged to $r>0.7R_{\rm P03}$. The profiles
are very well described by the NFW functional form and show an asymptotic inner
slope of $-1$. The deviation from NFW seen at $r\geq30~\rm{kpc}$ is due to the
presence of substructures in the haloes.

The situation is different for the haloes with baryonic physics (upper set of
curves). At $r\lesssim10~\rm{kpc}$, the profiles show significant differences
when the resolution is varied. Important differences are especially visible
between the two highest resolutions (blue and red curves). Clearly, the
criterion of \cite{Power2003} is no longer suitable because the profiles show
differences even at $r>R_{\rm P03}$ but note that in that range all three
resolutions nevertheless agree within $30\%$. Despite this poorer level of
convergence, the profiles display similar general trends across resolution
levels. The profiles are significantly steeper than NFW at $1.5~\rm{kpc}\lesssim
r \lesssim 10~\rm{kpc}$ and significantly shallower than NFW at smaller
radii. These differences relative to the NFW profile are larger than the
differences between the various resolution simulations, suggesting that the
trends seen are a generic outcome of the subgrid model assumed in our simulation
even if their exact magnitude is difficult to establish. In the remainder of
this paper, we will restrict our analysis to the highest resolution simulations,
but these limitations should be borne in mind when interpreting our
results and evaluating the generality of our conclusions.

\subsection{Sphericity of the distribution}

In order to characterise the morphology of the dark matter annihilation signal
at the centre of the Milky Way, it is interesting to study the shape of the dark
matter distribution. The profiles described so far assumed a spherically
symmetric dark matter density profile. With the higher precision of the
measurements of the excess and the increasing understanding of the GDE, it will
soon be possible to measure deviations from a perfect sphere. For instance, the
presence of a ``dark disc'' \citep{Read2008} would enhance the signal in the
plane of the galactic disk and hence break the symmetry of the signal. This
would also make the signal more difficult to disentangle from astrophysical
components associated with the disk, such as point sources.

In order to test this, we computed the inertia tensor of our four halos using
all the dark matter within a spherical aperture of $1~\rm{kpc}$ from the
centre. This distance corresponds to $\approx7\degree$ on the sky and hence
encompasses the majority of the $\gamma$-ray flux in the direction towards the
Galactic Centre that would result from dark matter annihilation in the MW. We
then compute the three eigenvalues $a>b>c$ of the inertia tensor and report the
values in Table \ref{tab:axisRatio} for both simulations with and without
baryons.

\begin{table}
\caption{Axis ratios ($a>b>c$) inferred from the inertia tensor of the matter
  within $1~\rm{kpc}$ from the centre of the galaxies for both the halos in the
  \dmo and \eagle local group simulations.}
\label{tab:axisRatio}
\begin{center}
\begin{tabular}{|c|c|c|c|c|}
     & \multicolumn{2}{c}{\dmo} & \multicolumn{2}{c}{\eagle} \\
Halo & $b/a$ & $c/b$ & $b/a$ & $c/b$\\
\hline 
$1$ & $0.888$ & $0.973 $ &  $0.989$ & $0.953 $ \\
$2$ & $0.863$ & $0.957 $ &  $0.975$ & $0.971 $ \\
$3$ & $0.850$ & $0.984 $ &  $0.981$ & $0.941 $ \\
$4$ & $0.879$ & $0.964 $ &  $0.987$ & $0.988 $ \\
\end{tabular}
\end{center}
\end{table}

As can be seen, the axis ratios are very close to unity, indicating only very
small deviations from sphericity and hence no obvious anisotropy feature in the
signal. We also find no alignment between the main axis of the dark matter
distribution in the inner $1~\rm{kpc}$ and the plane of rotation of the
stars. It is interesting to note that the simulation with baryons yields more
spherical distributions close to the centre than its counterpart without
baryons. We verified that repeating the exercise with apertures of $0.5$, $2$ and
$3~\rm{kpc}$ yields similar results.

\subsection{Position of the centre}

Another potential source of systematics in the analysis of the GC excess is the
position of the centre of the dark matter distribution. If the highest-density
part of the dark matter profile is offset from the centre of the stellar
distribution, then this offset should be detectable in observations. In their
simulation of a single Milky Way-like halo, \cite{Kuhlen2013} found a sizeable
offset of $300-400~\rm{pc}$ between the centre of the stellar distribution and
the peak of their dark matter distribution. If such an offset was indeed present
in the Milky Way, then an offset of $\approx2\degree$ between the GC and the
peak of the dark matter annihilation signal should be visible. In their study
based on the \eagle simulations, \cite{Schaller2015a} found that the offset
between the peak of the dark matter density distribution and the centre of the
stellar distribution is typically smaller than the softening length of the
simulation ($\epsilon = 700~\rm{pc}$ in their case). Repeating their analysis on
our four simulated Milky Way halos, we find offsets between $22$ and
$43~\rm{pc}$, well below the size of the softening length
($\epsilon=134~\rm{pc}$), indicating that the offsets are consistent with
zero. For all practical purposes and given the current resolution of
instruments, the centre of the dark matter distribution is hence coincident with
the centre of the stellar distribution.

\section{Dark matter annihilation signal}
\label{sec:signal}

Now that the dark matter profiles have been characterised, we turn to the
derivation of the corresponding annihilation signal as observed by a virtual
instrument located at the position of the Sun and pointing towards the centre of
the Milky Way.

\subsection{J-factor for the simulated halos}

In the case of a dark matter particle that annihilates into photons or into
particles that generate photons in their decays, the photon flux (in units of
${\rm GeV^{-1}\cdot cm^{-2} \cdot s^{-1} \cdot sr^{-1}}$) at a given angle,
$\Psi$, on the sky away from the GC is given by

\begin{equation}
 \frac{{\rm d}N}{{\rm d}E}(\Psi)= \frac{\langle\sigma v \rangle}{8\upi 
m_\chi^2}\frac{{\rm d}N_\gamma}{{\rm d}E}I(\Psi),
\label{eq:flux}
\end{equation}
where $m_\chi$ is the mass of the dark matter particle, $\langle\sigma v\rangle$ is
its velocity averaged total annihilation cross section, ${\rm d}N_\gamma/{\rm d}E$ is
the averaged energy spectrum of photons produced per annihilation and $I(\Psi)$ is
the integral along the line of sight of the square of the dark matter density. This
so-called ``J-factor'' reads

\begin{equation}
 I(\Psi) = \int_{\rm l.o.s.} \rho_{\rm DM}^2(r(s,\Psi))~{\rm d}s,
\label{eq:Jfactor}
\end{equation}
with the variable $s$ running along the line of sight axis from $s=0$ to $s=\infty$ and 
\begin{equation}
 r(s,\Psi) = \sqrt{(\rsun-s\cos\Psi)^2+(s\sin \Psi)^2}
\end{equation}
giving the distance from the GC for a particular angle on the sky
$\Psi$ and distance to the GC, $\rsun$. The differential intensity
$dN/dE$ is hence the product of the J-factor, given by the
distribution of dark matter, and the particle physics model assumed.
As a consequence, within a reasonable range, the precise normalisation
of the J-factor is irrelevant since a similar signal can be recovered
by altering the particle physics model. To simplify the comparison with
the analysis of the GC excess, we have, thus, normalised our simulated
profiles such that $\rho_{\rm DM}(\rsun) =
0.4~\rm{GeV}\cdot\rm{cm}^{-3}$.

\begin{figure*}
\includegraphics[width=1.\textwidth]{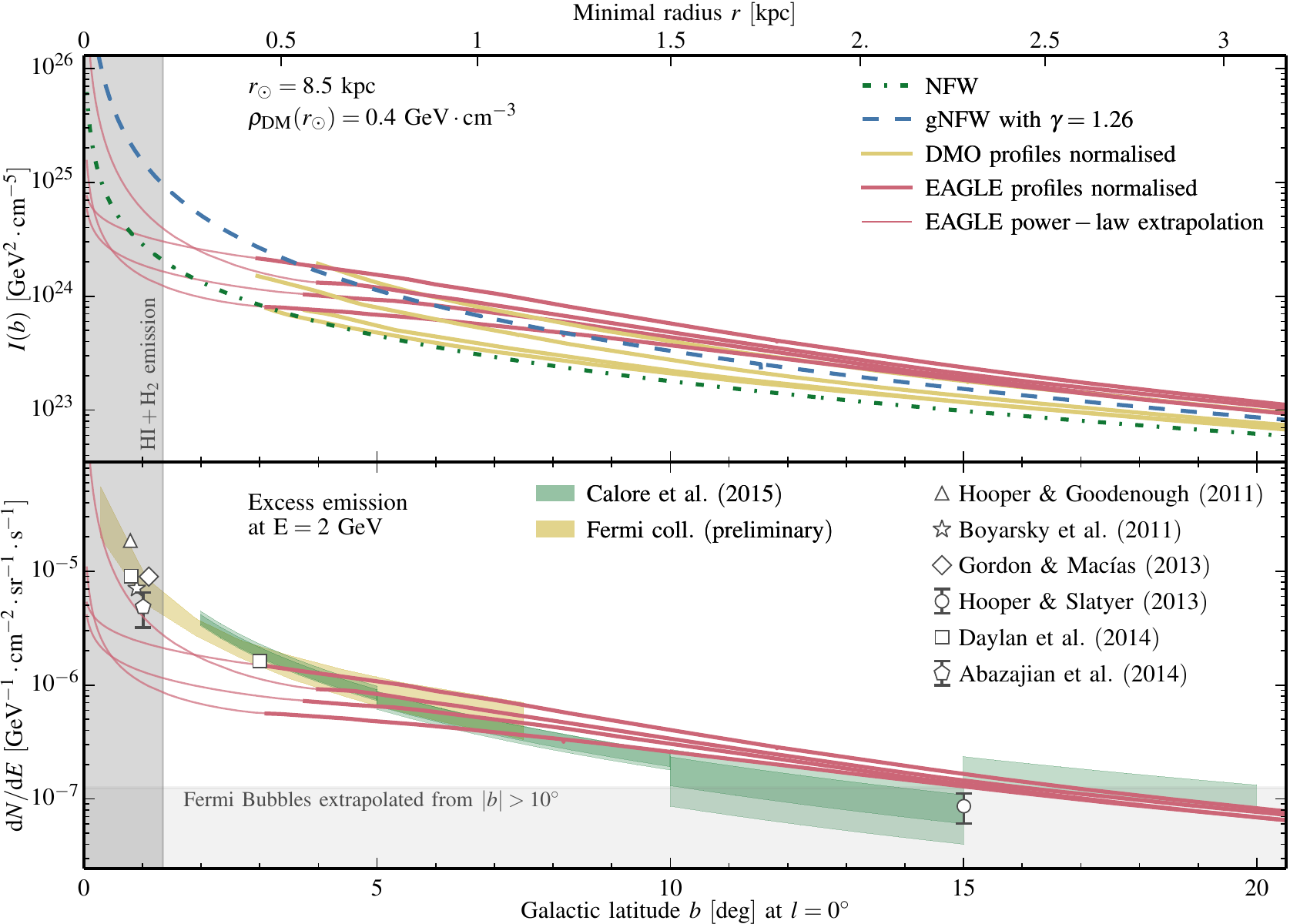}
\caption{{\it Top panel:} The J-factor as a function of galactic latitude $b$
  inferred from our four simulated halos both with (red lines) and without
  (yellow lines) baryon effects. The green dot-dashed and blue dashed lines
  correspond to NFW and gNFW with $\gamma=1.26$ profiles respectively with a
  scale length $r_{\rm s}=20~\rm{kpc}$. To ease comparison, all profiles have
  been normalised to yield $\rho_{\rm
    DM}(\rsun)=0.4~\rm{GeV}\cdot\rm{cm}^{-3}$. The thin lines correspond to
  power-law extrapolations of our simulated profiles (see text). The scale at
  the top indicates the minimum radius intersected by a line of sight at the
  given galactic latitude $b$.  {\it Bottom panel:} Emission at $E=2~{\rm GeV}$
  for our halos assuming the best-fitting particle physics model from
  \citet{Calore2015b}. Data points with error bars show the best-fitting models
  of \citet{Hooper2010}, \citet{Boyarsky2011}, \citet{Gordon2013},
  \citet{Hooper2013}, \citet{Daylan2014} and \citet{Abazajian2014} to the Fermi
  GeV excess. The green shaded regions indicate the excess emission and its
  statistical uncertainty for a fixed gNFW profile derived by
  \citet{Calore2015a} and the yellow shaded region indicates the preliminary
  results of the Fermi-LAT team. The vertical grey shaded region indicates the
  radial range where uncertainties in the GDE modelling due to $\pi^0$ emission
  from \ion{H}{I} and $\rm{H}_2$ regions dominate the foreground templates used
  in the analysis of the data \citep[adapted from][]{Calore2015b}. Similarly,
  the shaded region at the bottom indicates the flux intensity of the Fermi
  bubbles. }
\label{fig:flux}
\end{figure*}

In the top panel of Fig.~\ref{fig:flux} we show the J-factor
(Eq.~\ref{eq:Jfactor}) as a function of galactic latitude $b$ (at galactic
longitude $l=0\degree$) for our four simulated profiles, normalised to the same
local dark matter density. The red and yellow lines correspond to the the dark
matter profiles in the simulations with and without baryons respectively. The
green dot-dashed and blue dashed lines correspond to NFW and gNFW with
$\gamma=1.26$ profiles respectively with a scale length $r_{\rm s}=20~\rm{kpc}$
and the same normalisation local dark matter density as our normalised
halos. The shape of the J-factor profile is different in the runs with and
without baryons. The contraction of the dark matter due to baryons increases the
J-factor by a factor of $\approx 2$ at angles $b\gtrsim4\degree$ from the GC,
when compared to an NFW halo. In that angular range, the J-factor is also larger
than the one obtained for a gNFW with a slope $\gamma=1.26$
\citep{Calore2015a}. Closer to the GC, the simulated J-factors display a
shallower slope and values lower than the gNFW model.

\subsection{Extrapolation of the profiles towards the centre}

As most of the dark matter annihilation signal originates from the inner few
hectoparsecs, it is necessary to extrapolate our findings from section
\ref{ssec:profiles} to smaller radii. As the annihilation signal increases with the
square of the local density, one can ask what the highest density is that can be
reached in the inner regions given the constraints on the density and enclosed mass
at the smallest converged radius. Assuming that the profile is not hollow and that
the logarithmic slope is monotonic, it is straightforward to show that the only
asymptotic power law that can be used to extrapolate the profiles from a given radius
$r$ towards the centre has a slope $\gamma_{\rm max} = 3(1-4\upi r^3\rho(r)/3M(r))$
\citep{Navarro2010}. Setting $r$ to the convergence radius of the halos, $R_{\rm
  P03}$, we obtain slopes in the range $\gamma_{\rm max}=0.55-1.22$ for our four
halos. The J-factors resulting from these extrapolations are shown on
Fig.~\ref{fig:flux} using thin lines. They allow us to set upper bounds on the
J-factor for angles $b\lesssim3\degree$. Even with this power-law extrapolation, the
flux is lower than the gNFW profile with a slope $\gamma=1.26$.

\subsection{Gamma-ray flux morphology}

In the bottom panel of Fig.~\ref{fig:flux}, we show the emission at
$E=2~\rm{GeV}$ for our J-factors, assuming the best-fitting particle physics
model of \cite{Calore2015a}\footnote{$m_\chi=46.6~\rm{GeV}$, $\langle\sigma
  v\rangle=1.60\times10^{-26}~\rm{cm}^3~\rm{s}^{-1}$ for a $b\bar b$ final
  annihilation state.}. For comparison, we show the flux inferred from the GC
excess by \cite{Hooper2010},\cite{Boyarsky2011}, \cite{Gordon2013},
\cite{Hooper2013}, \cite{Daylan2014} and \cite{Abazajian2014} with the error
bars indicating the $\pm1\sigma$ statistical error (not shown when smaller than
the symbols). The observed intensities were rescaled following the procedure
highlighted in \cite{Calore2015b}, taking into account the assumed excess
profiles. Note that individual measurements are more than $3\sigma$ discrepant
with each other. The green shaded regions indicate the best-fitting model of
\cite{Calore2015a}.  Their model assumed a gNFW profile for the dark matter
profile and used $60$ GDE templates in their likelihood analysis of $10$ regions
of interest on the sky located around the galactic centre. The width of the
green regions on the figure indicates both the statistical uncertainty and the
posterior range of the GDE modelling around the best-fitting gNFW profile. The
uncertainty on the slope of the profile is not shown. A similar analysis,
performed by \cite{Calore2015b}, of the preliminary results of the Fermi
collaboration is shown as a yellow shaded region. The grey shaded region on the
left of the plot indicates the radial range over which the emission from the ISM
gas dominates the GDE models \citep{Calore2015b}.  Similarly, the grey shaded
region at the bottom of the panel indicates the level of $\gamma$-ray flux
expected from the extended ``Fermi Bubbles'' \citep{Su2010}, thought to be the
remnant of past AGN activity. We use the extrapolation, assuming a constant
density, to lower latitudes of the flux estimated by \cite{Fermi2014}. The flux
originating from the annihilation of dark matter is higher than the contribution
of the Fermi Bubbles at angles $b<15\degree$, making the radial range
$2\degree<b<15\degree$ ideal for the study of the excess
\citep{Calore2015b}. The resolution of our simulations is, hence, well matched
to this requirement.

Our simulated profiles (red lines) are in good agreement with the $\gamma$-ray
data for angles $b>3\degree$. This is expected since over the relevant radial
range, the profiles have a similar form to the gNFW profile with asymptotic
inner slope, $\gamma=1.2-1.3$, inferred directly from the data (assuming that
the emission is due to dark matter annihilation).  At smaller angles, three of
the four extrapolated profiles give significantly less emission than observed,
whilst the fourth is in good agreement with the data. We stress, however, that
this extrapolation gives the largest power-law signal at the GC. The predicted
emission at $b<2\degree$ could be boosted by adjusting the particle physics
model but there is a danger that such adjustments could lead to an
overprediction of the emission at larger angles.

\section{Summary \& Conclusion}
\label{sec:summary}

In this study we investigated the dark matter density profiles of four
Milky Way galaxies simulated using a state-of-the-art hydrodynamics
code and subgrid model. We specifically focused on the inner few
kiloparsecs of the dark matter distribution in order to refine earlier
predictions for the properties of dark matter annihilation emission.
The careful treatment of baryons in our simulations allows us to
understand and analyse the effects that baryons have on the dark
matter distribution. These are not negligible and give rise to haloes
whose properties differ significantly from those of the ``standard
halo model'' often used in dark matter detection studies. Whilst our
simulations are among the highest resolution examples of their kind
currently available, we are only able to establish convergence within
a tolerance of $30\%$ across different resolutions over the radial
range of interest (sec.~\ref{sec:convergence}). We feel that this
level of convergence is sufficient to support our conclusions but
higher resolution simulations will be needed to test this
supposition. 

We can summarise our findings as follows: 

\begin{itemize}

\item As seen in previous simulations  \citep[e.g.][]{Dubinski1994, Abadi2010, Bryan2013},
  the central concentration of baryons significantly reduces the asphericity
  typical of haloes in dark matter-only simulations. The distribution of dark
  matter in the inner $500~\rm{pc}$ is very close to spherical with axis ratios,
  $b/a>0.96$, in all cases. \\

\item There is no detectable offset between the position of the GC and the peak
  of the dark matter distribution. The largest offset found in our halos is
  $45~\rm{pc}$, much smaller than the softening length of the simulations
  ($\epsilon=134~\rm{pc}$). \\

\item The condensation of baryons at the halo centre causes the halo to contract
  slightly. The halo density profiles end up having steeper profiles than NFW in
  the radial range $r=2-10~\rm{kpc}$. \\

\item In the inner $1.5~\rm{kpc}$, which are well resolved in our
  simulations the dark matter halo density profiles develop
  significant flattening. This feature is likely to be associated with
  violent star formation events that take place during the early
  stages of galaxy formation. It must be borne in mind that effects of
  this kind are sensitive to the specific subgrid physics model and,
  at this point, they must not be regarded as a generic prediction of the \lcdm model.\\

\item The predicted dark matter annihilation emission signal is in
  good agreement with the detection of extended $\gamma$-ray emission
  in excess of the known foregrounds by the \emph{Fermi} satellite at
  galactic latitudes, $b\gtrsim3\degree$, where our haloes are well
  resolved. A simple extrapolation of the density profiles in our
  simulations to smaller angles predicts a $\gamma$-ray flux
  significantly lower than is measured, in three of the four cases,
  suggesting possible contributions from other sources to the excess.
  In the fourth case, the annihilation signal from the extrapolation
  is in broad agreement with the reported measurements close to the
  Galactic Centre.
\end{itemize}

The analysis of the \emph{Fermi} excess has so far been performed assuming a
gNFW profile or other parametric profile forms for the dark matter. Future, more
precise studies, would benefit from using the more realistic profile shapes
derived directly from hydrodynamical simulations. This should help disentangle
the signal from dark matter annihilation from the galactic diffuse emission.

\section*{Acknowledgements}
This work would have not be possible without Lydia Heck and Peter Draper's
technical support and expertise.  This work was supported by the Science and
Technology Facilities Council (grant number ST/F001166/1); European Research
Council (grant numbers GA 267291 ``Cosmiway'' and GA 278594
``GasAroundGalaxies'') and by the Interuniversity Attraction Poles Programme
initiated by the Belgian Science Policy Office (AP P7/08 CHARM). RAC is a Royal
Society University Research Fellow. \\  This work used the DiRAC Data Centric
system at Durham University, operated by the Institute for Computational
Cosmology on behalf of the STFC DiRAC HPC Facility (www.dirac.ac.uk). This
equipment was funded by BIS National E-infrastructure capital grant
ST/K00042X/1, STFC capital grant ST/H008519/1, and STFC DiRAC Operations grant
ST/K003267/1 and Durham University. DiRAC is part of the National
E-Infrastructure.  We acknowledge PRACE for awarding us access to the Curie
machine based in France at TGCC, CEA, Bruy\`eres-le-Ch\^atel.

\bibliographystyle{mnras} 
\bibliography{./bibliography.bib}

\appendix

\section{Resolution test for all haloes}
\label{sec:allConvergence}

\begin{figure*}

\includegraphics[width=\columnwidth]{Figures/DMprofile_resolution_1.pdf}
\includegraphics[width=\columnwidth]{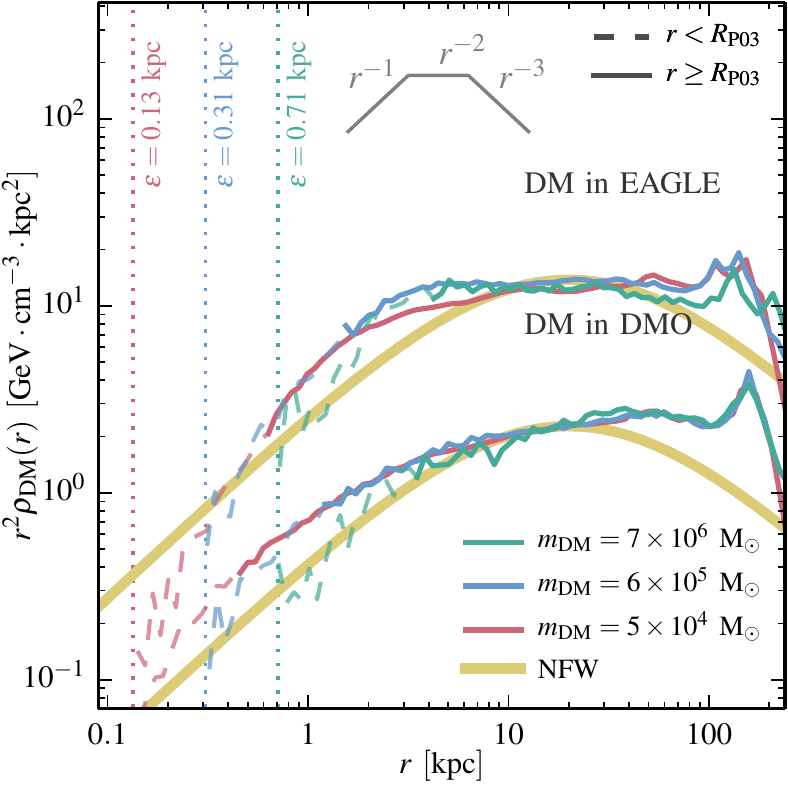}
\includegraphics[width=\columnwidth]{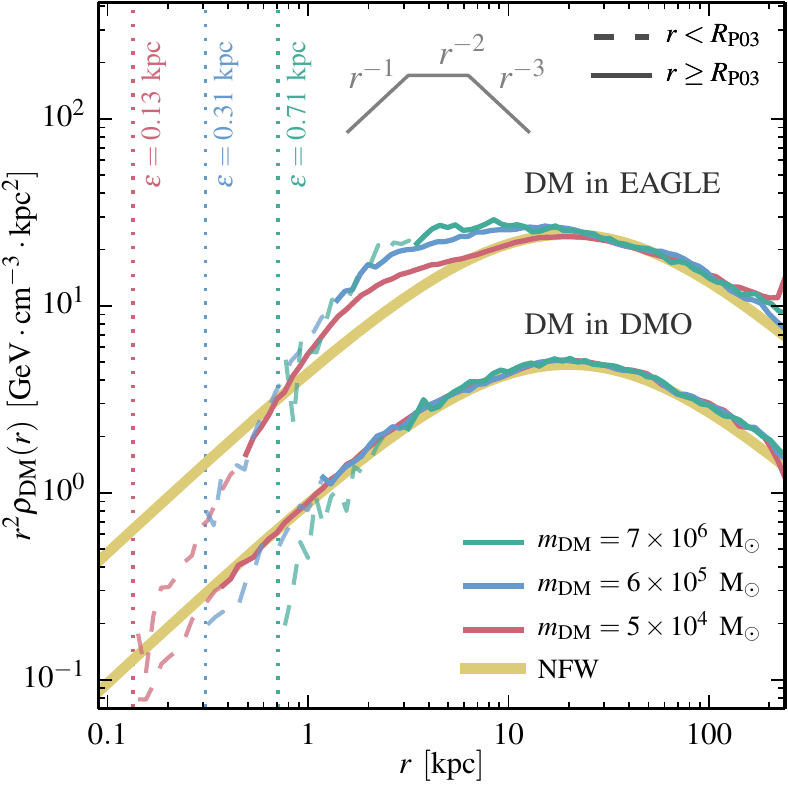}
\includegraphics[width=\columnwidth]{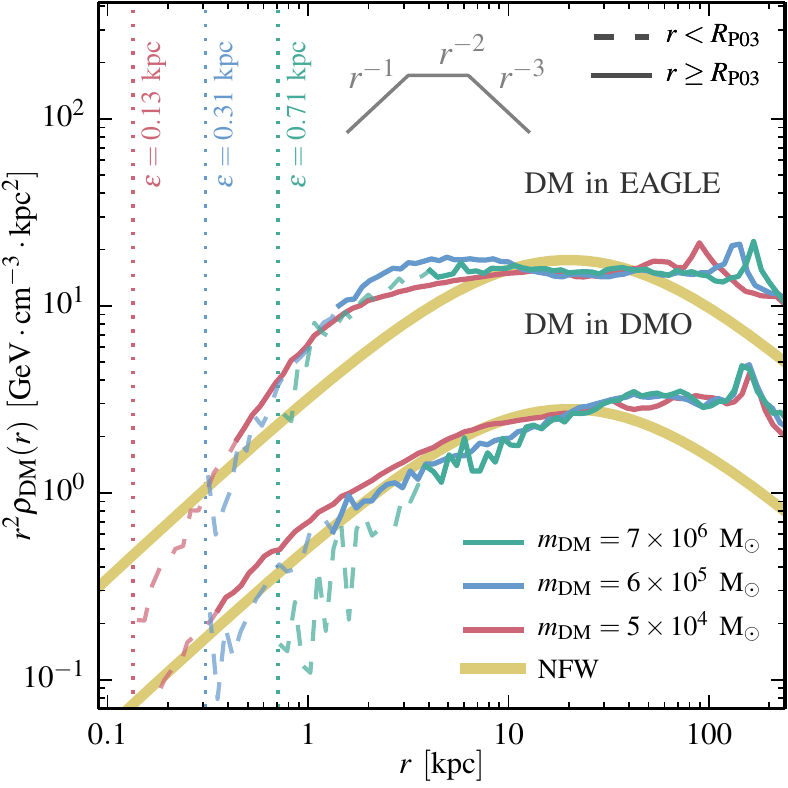}
\caption{Same as Fig.~\ref{fig:convergence} but for all four halos
  considered in this study. The DM density profiles in all four haloes agree
  to better than $30\%$ at $r<10~\rm{kpc}$.}
\label{fig:allConvergence}
\end{figure*}

\label{lastpage}

\end{document}